\documentclass[a4paper]{article}

\usepackage{INTERSPEECH2022}
\usepackage[T1]{fontenc}
\usepackage{caption, subcaption}
\usepackage[justification=centering]{caption}
\usepackage{adjustbox}
\usepackage{mathtools,xparse}
\usepackage{amsmath,amssymb,fixmath}
\usepackage{graphicx}
\usepackage{float}
\usepackage{cite}
\usepackage[dvips]{epsfig}
\usepackage{multirow, multicol}
\usepackage{makecell}
\usepackage{multicol}
\usepackage{tabularx}
\usepackage{hhline}
\usepackage{arydshln}
\usepackage{nicefrac}
\usepackage{color}
\usepackage{verbatim}
\usepackage{tablefootnote}
\usepackage{txfonts}
\usepackage[usenames,dvipsnames]{xcolor}
\usepackage{enumitem}
\usepackage{hyperref}

\newcommand{\yedit}[1]{\textcolor{black}{#1}}
\newcommand{\oedit}[1]{\textcolor{black}{#1}}
\newcommand{\eedit}[1]{\textcolor{black}{#1}}

\title{Language Model-Based Emotion Prediction Methods for Emotional Speech Synthesis Systems}
\name{Hyun-Wook Yoon$^1$, Ohsung Kwon$^1$, Hoyeon Lee$^1$, Ryuichi Yamamoto$^2$,\\Eunwoo Song$^1$, Jae-Min Kim$^1$, and Min-Jae Hwang$^1$}
\address{
  $^1$NAVER Corp., Seongnam, Korea, $^2$LINE Corp., Tokyo, Japan}
\email{hyunwook.yoon@navercorp.com}

\begin{document}

\fontsize{8.5}{9.35}\selectfont 

\maketitle
\begin{abstract}
This paper proposes an effective emotional text-to-speech (TTS) system with a pre-trained language model (LM)-based emotion prediction method.
Unlike conventional systems that require auxiliary inputs such as manually defined emotion classes, our system directly estimates emotion-related attributes from the input text.
Specifically, we utilize generative pre-trained transformer (GPT)-3 to jointly predict both an emotion class and its strength in representing emotions’ coarse and fine properties, respectively.
Then, these attributes are combined in the emotional embedding space and used as conditional features of \oedit{the} TTS model for generating output speech signal\oedit{s}.
Consequently, the proposed system can produce emotional speech only from text without any auxiliary inputs.
Furthermore, because the \eedit{GPT-3} enables to capture emotional context among the consecutive sentences, the proposed method can effectively handle the paragraph-level generation of emotional speech.
\end{abstract}

\section{Introduction}
As the quality of the neural text-to-speech (TTS) system has reached natural sound almost indistinguishable from human recordings~\cite{jonathan2017natural, lee2020multi, ren2020fastspeech2}, the interest in emotional TTS \yedit{has \eedit{also} increased}~\cite{kwon2019effective,cai2021emotion,lu2021multi,wang2018style,wu2019end}.
One of the general approaches is using reference audio~\cite{wang2018style, henter2018deep, wu2019end, zhang2019learning}, \eedit{where} the emotional style representation, i.e., emotion embedding, is extracted from the emotional reference audio and then used as an auxiliary input of the TTS system.

Directly utilizing \eedit{natural speech's emotion information} as the conditional input can effectively deliver the expressiveness to synthetic outputs.
However, the overall cost \oedit{of} deciding \eedit{the} proper emotional attributes in each sentence can be high since these methods require \oedit{a} continuous selection process of auxiliary inputs during an inference.
For instance, when synthesizing a bulk of sentences, such as \yedit{audiobook contents}~\cite{cambre2020choice}, the entire TTS process becomes expensive since a large number of emotions \yedit{need} to be annotated properly before synthesizing the corresponding waveforms.

To address \oedit{the} aforementioned problem, we propose an language model (LM)-based \eedit{automatic} emotion prediction method.
\eedit{From the text inputs}, a \textbf{\textit{joint emotion predictor}} estimates the emotion class and its strength to model the coarse and fine properties of the target emotion, respectively.
Specifically, we adopt a generative pre\oedit{-}trained transformer (GPT)-3~\cite{brown2020language} as the backbone network of \yedit{the} emotion predictor that enables to predict the emotional attributes without any auxiliary information.
Then, a \textbf{\textit{joint emotion encoder}} \eedit{converts} the emotion class and strength \eedit{into the} embedding space by linearly scaling the embedding vector of the emotion class according to its strength.
Finally, the \eedit{resulting} embedding is fed to the TTS model to synthesize emotional output speech.
Our contributions are summarized as follows:
\begin{itemize}[leftmargin=*]
    \item 
        To achieve \yedit{the} emotional TTS system without relying on auxiliary inputs, we propose an LM-based emotion prediction method that estimates both the emotion class and strength of the target speech from input text.
        The experimental results showed that the proposed method estimated more accurate emotional attributes than the conventional method.
    \item 
        We tested the proposed emotion prediction method in a neural TTS system based on a Tacotron 2 acoustic model~\cite{okamoto2019tacotron, song2020neural, hwang2021tts} with a Parallel WaveGAN (PWG) vocoder~\cite{yamamoto2020parallel, song2021improved, yamamoto2021parallel, hwang2021high}.
        The experimental results verified that the proposed TTS system synthesized more natural and expressive speech than the conventional method, and was competitive with the system that used manually defined emotional attributes.
        In particular, our TTS system achieved results of 3.92 and 3.94 in the mean opinion score (MOS) tests, which evaluated the naturalness and emotional expressiveness of synthetic speech.        
    \item
        Under the paragraph-level speech synthesis scenario, we showed that the proposed system was capable of replacing the emotional TTS system with human-annotated emotional attributes.
        A subjective preference test verified that these TTS systems were perceptually indistinguishable.
\end{itemize}

\section{Related Works}
\label{sec:relatedwork}
\eedit{In the emotional TTS tasks}, several studies have proposed \eedit{to predict} emotional attributes from the text \eedit{input}.
For instance, Cui et al.~\cite{cui2021emovie} proposed \eedit{to predict} an emotion class from \oedit{the} text by using a convolutional neural network (CNN) block.
Lei et al.~\cite{lei2021fine} proposed \eedit{to predict} emotion strength from \oedit{the} text by using the text encoder module of Tacotron~\cite{wang2017tacotron}.
Although these methods provided highly emotional speech, their applications were limited because they required manual definitions of the desired emotion class~\cite{cui2021emovie} or strength~\cite{lei2021fine} during the inference phase, which makes the cost of generating emotional speech much higher.

Unlike those studies, our method differs in that it predicts both the emotion class and the emotion strength by using pre-trained LM.
\eedit{Note that} Lei et al.~\cite{lei2022msemotts} adopted a pre-trained BERT model~\cite{devlin2018bert} to predict emotional attributes from text, which is mostly similar to our work.
However, the difference \oedit{in} our work is that we adopt GPT-3 instead of BERT, and further propose a joint emotion encoder that maximizes the merits in the usage of emotion strength.

\section{Proposed Emotional TTS System}
\label{sec:proposed}

\begin{figure}[t!]
	\centering
	\includegraphics[width=1.0\linewidth]{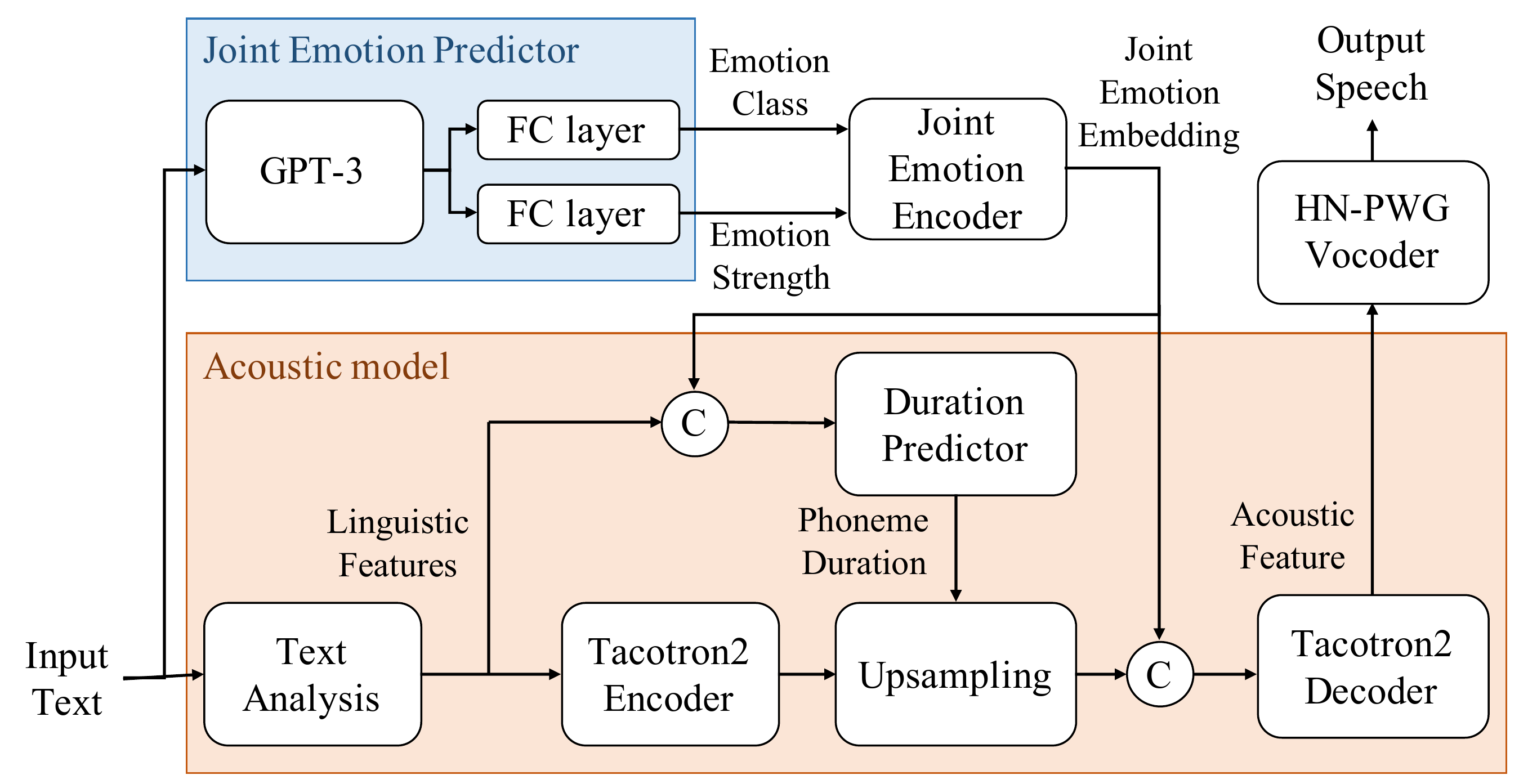}
	\caption{
	    Block diagram of the proposed system;
        $\copyright$ denotes the concatenation process between input tensors.
	}
	\label{fig:prop_system}
	\vspace*{-6pt}
\end{figure}

The proposed emotional TTS system is illustrated in Figure~\ref{fig:prop_system}.
First, the joint emotion predictor estimates the emotion class and strength from the input text.
Then, the joint emotion encoder mixes them to reflect the joint representation of emotions’ coarse and fine characteristics. 
Finally, this joint emotion embedding is fed to the TTS model to synthesize output emotional speech.
The details of this process are described in the following sections.

\subsection{Emotion class and strength annotation}
\label{sec:annotation}
Before training the joint emotion predictor, it is necessary to prepare the annotations of emotion class and strength from the recorded data.
In the case of the emotion class annotation, we guide the speaker to act \oedit{on} the specific emotion during the recording process by assuming the speaker's acting is enough to appropriately represent the target emotion.

However, annotating the emotion strength is a relatively difficult task, since its subjective and relative properties make it hard to quantify.
To overcome this problem, we adopt an approach similar to Lei et al.~\cite{zhu2019controlling} based on a ranking support vector machine (RankSVM)~\cite{parikh2011relative}, which is a well-known method for measuring relative attributes between binary classes.
Specifically, the emotion features consisting of Mel-frequency cepstral coefficients (MFCC), fundamental frequency (F0), and so on are first extracted from the neutral and emotional speech.
Then, a ranking function between these features is trained with the RankSVM objective~\cite{joachims2002optimizing, parikh2011relative}.
After computing a rank for \oedit{the emotion of each speech signal}, it is normalized to have a value between 0 (weakest) and 1 (strongest); we then define it as annotated emotion strength.
Note that, based on the assumption that a neutral emotion has no strength, we fixed the strength of neutral emotion at zero.
Consequently, it is possible to quantify the relative strength that represents the intensity of emotion compared to the neutral data.

\subsection{Emotion class and strength prediction from text}
To construct an efficient emotional TTS system, we propose a \textbf{\textit{joint emotion predictor}} that estimates both the emotion class and its strength from input text.
To accurately predict these attributes, we utilize GPT-3 as a feature extractor network of \yedit{the} emotion predictor \cite{brown2020language, kim2021changes}.
In particular, the proposed emotion predictor consists of the GPT-3 followed by two individual fully connected (FC) layers, where one FC layer is used to estimate the one hot encoded emotion class and the other is used to estimate the scalar-valued emotion strength.

The advantages of the proposed emotion predictor are clear.
First, \oedit{due to} the GPT-3 enables the model to accurately predict emotional attributes, it is capable of getting rid of the \oedit{handwork for emotion annotation of the speech signal} which makes the emotional TTS expensive.
Additionally, thanks to the GPT-3's ability to \oedit{identify} the contextual information of input sentences, it effectively considers the emotional context among multiple sentences.

\subsection{Joint embedding of emotion class and strength}
\label{sec:emotion_encoder}
Once the emotion's class and strength are generated, the \textit{\textbf{joint emotion encoder}} blends them to compose their joint emotion embedding \oedit{vector}.
First, the one-hot encoded emotion class is converted to a corresponding emotion class embedding through a look-up table (LUT).
Then, the joint emotion embedding, $\mathbf{h}_e$, is obtained by mixing emotion class embedding, $\mathbf{e}_{emb}$, and the emotion strength, $e_{str}$, as follows:
\begin{equation}
    \label{eq:emo_enc}
    \mathbf{h}_e = \text{softplus}\left( \mathbf{W}_{emb} \mathbf{e}_{emb} \cdot \left(1 + w_{str}e_{str} \right) \right),
\end{equation}
where $\mathbf{W}_{emb}$ and $w_{str}$ denote the projection matrix and scalar variable, respectively; 
$\text{softplus}( \cdot )$ denotes the softplus activation function~\cite{zheng2015improving}. 
Finally, the joint emotion embedding is fed to the TTS model for the synthesis of emotional speech.

\begin{figure}[t!]
	\centering
    \includegraphics[width=0.8\linewidth]{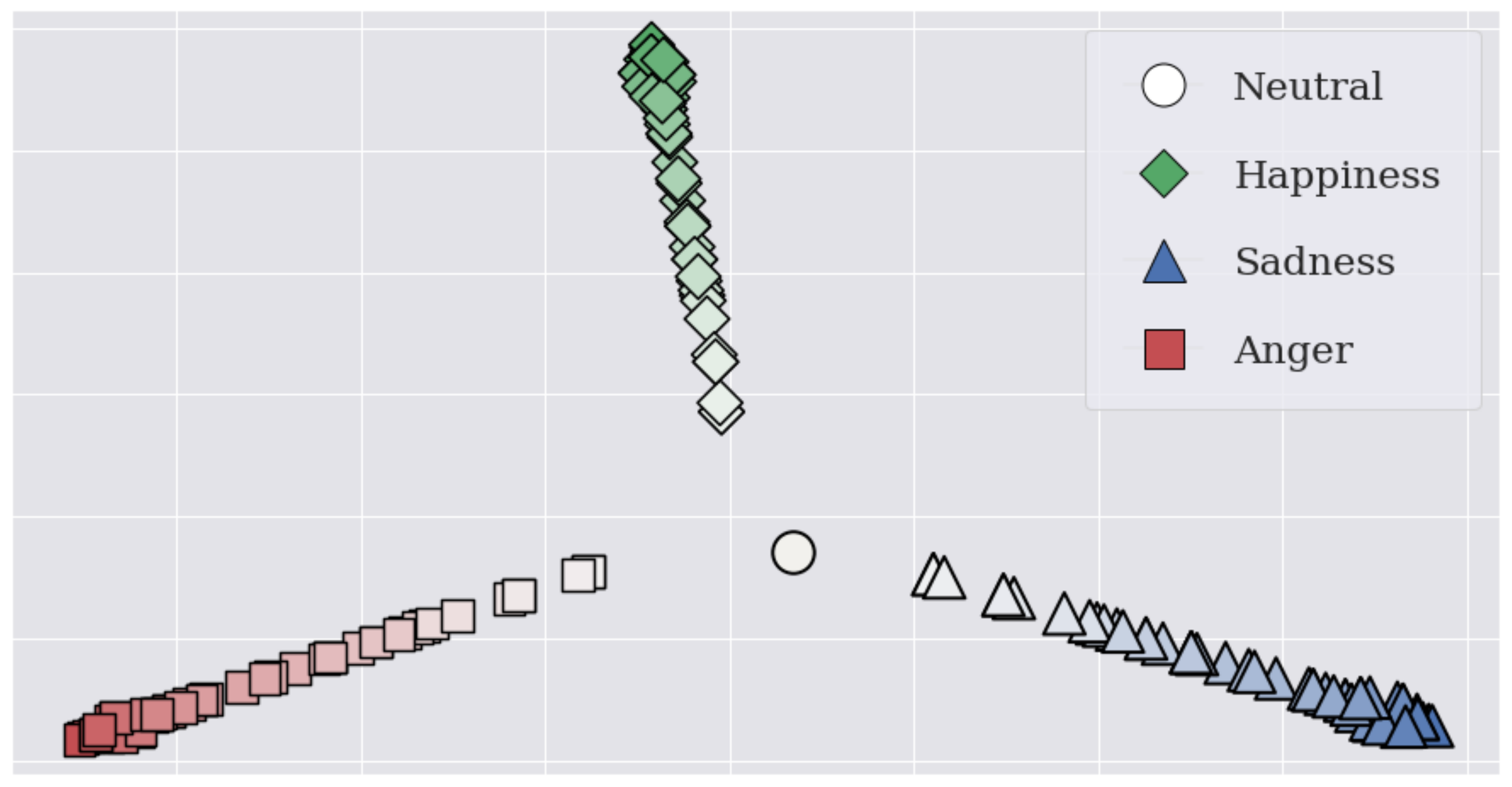}
	\caption{
	    t-SNE plot of joint emotion embeddings generated from randomly sampled emotion strengths with various emotions.
        The brighter color indicates that the emotion strength becomes stronger, and vice versa.
    }
	\label{fig:t_sne_plot}
	\vspace*{-6pt}
\end{figure}

By utilizing both emotion class and strength, the proposed method can effectively reflect both the coarse and fine structures of emotion properties.
To show this, we drew the t-distributed stochastic neighbor embedding (t-SNE)~\cite{van2008visualizing} plot of joint emotion embedding, $\mathbf{h}_e$, which is illustrated in Figure~\ref{fig:t_sne_plot}. 
The result shows that each joint emotion embedding converges to the single vector point, which corresponds to the embedding of neutral emotion as the emotion strength weakens; whereas it diverges in a particular direction as the emotion strength increases.
This indicates that each emotion class compose\yedit{s} its own cluster in the emotional embedding space, which corresponds to the emotion's coarse property; whereas emotion strength can adjust the emotion embedding to finely represent the emotional properties.

\subsection{TTS model}
\label{sec:tts_model}
As an acoustic model of \yedit{our} TTS system, we adopt a duration-informed Tacotron 2~\cite{okamoto2019tacotron, song2020neural, hwang2021tts}, which has the capacity to accurately align the phoneme sequence with the acoustic features.
In this method, the linguistic features extracted from input text are fed to the Tacotron 2 encoder to extract high-level linguistic embeddings.
Then, the external duration predictor estimates emotion-dependent phoneme duration by receiving the concatenated vectors of linguistic features and joint emotion embeddings in Eq.~\eqref{eq:emo_enc} as inputs.
For given phoneme durations, the high-level linguistic embedding is up-sampled through a repetition process to match its temporal resolution to that of the acoustic features.
Finally, the Tacotron 2 decoder \cite{jonathan2017natural} autoregressively generates the output acoustic features by receiving concatenated vectors of up-sampled linguistic embedding and joint emotion embedding.

After generating acoustic features, the multi-band harmonic-plus-noise PWG (MB-HN-PWG) vocoder~\cite{hwang2021high} synthesizes the output speech.
In this vocoder, two separated harmonic and noise WaveNets jointly capture the harmonic and noise components of speech within the PWG framework to generate high-quality speech waveforms.
As a result, it can stably generate qualified speech signal\oedit{s} with desired emotion.

\subsection{Training criteria}
Our models consist of two \yedit{separated parts}: the \yedit{joint} emotion predictor and the acoustic model.
The loss function of each \yedit{part} is described as follows:
\begin{gather}
    \mathcal{L}_{a} = \mathcal{L}_2(\mathbf{\hat{y}}, \mathbf{y}) + \mathcal{L}_2(\mathbf{\hat{d}}, \mathbf{d})\\
    \label{eq:emo_loss}
    \mathcal{L}_{e}= \mathcal{L}_2(\hat{e}_{str}, e_{str}) + \lambda_{cls} \cdot CE(\mathbf{\hat{e}}_{cls}, \mathbf{e}_{cls}),
\end{gather}
where $\mathcal{L}_{a}$\footnote{
    During a training stage, the target emotion class and strength are used in the TTS model instead of predicted values to avoid the effect of \oedit{the} prediction error of emotion predictor.
} and $\mathcal{L}_{e}$\footnote{
    For the computational efficiency, we froze the parameters of the GPT-3 during the TTS's model training.
} denote the loss functions of the acoustic model and emotion predictor, respectively;
$\mathbf{y}$, $\mathbf{d}$, $\mathbf{\hat{y}}$, and $\mathbf{\hat{d}}$ denote acoustic features, phoneme duration, and their generated counterparts, respectively;
$\mathcal{L}_2( \cdot, \cdot )$ and $CE( \cdot, \cdot )$ denote L-2 loss and cross entropy loss, respectively;
$\lambda_{cls}$ denotes the hyperparameter balancing the two loss terms, which was chosen to be 0.01 based on our preliminary experiments.

\section{Experiments}
\subsection{Database and features}

For the experiments, a phonetically balanced emotional TTS corpus recorded by a female Korean professional speaker was used.
The speech corpus consisted of four different emotions: neutral, happiness, sadness, and anger.
Specifically, 2,000, 100, and 30 utterances were used as each emotion's training, validation, and test sets, respectively.
In total, 8,000 utterances (8.72~h), 400 utterances (0.43~h), and 120 utterances (0.16~h) were used for the training, validation, and test sets, respectively.
The speech signals were sampled at 24-kHz with 16-bit quantization.

As the input of the TTS acoustic model, we used 354-dimensional linguistic feature \oedit{vectors}.
Specifically, 330-dimensional \oedit{vectors} including phoneme, accent, break, and punctuation were used as categorical linguistic features; whereas 24-dimensional \oedit{vectors} including positional information of each phoneme were used as numerical features.

As the output vectors of the TTS acoustic model, we used 79-dimensional feature \oedit{vectors} extracted by an improved time-frequency trajectory excitation vocoder~\cite{song2017effective}, which consisted of 40-dimensional line spectral frequencies, F0, energy, binary voicing flag, a 32-dimensional slowly evolving waveform, and a 4-dimensional rapidly evolving waveform.
The frame length and the shifting size were set to 20 ms and 5 ms, respectively.

As emotion features for training rankSVM, the 384-dim feature \oedit{vectors were} extracted from the speech signal by using the openSMILE tooklit~\cite{eyben2013recent}, which consisted of a 12-dimensional MFCC, harmonic-to-noise ratio, root mean square frame energy, zero-crossing rate, and F0 with their statistical parameters~\cite{schuller09_interspeech}.

\subsection{Model configuration}
\subsubsection{Emotion predictor and encoder}
\label{sec:emotion_predictor}
As a baseline emotion predictor, we adopted a system proposed in Cui et al.~\cite{cui2021emovie}, which predicted emotion class by using a 2-dimensional CNN model.
In particular, the baseline emotion predictor consisted of three convolution blocks composed of two 2-dimensional convolution layers with a 3$\times$3 kernel size followed by a 2-dimensional max pooling operation layer.
In each convolution block, the input/output dimensions were set to 1/64, 64/128, and 128/256, respectively.
Finally, two FC layers with output dimensions of 128 and four were used as the last layers for predicting the emotion class.
After predicting the emotion class, emotion embedding was composed by passing it to the LUT layer.

For the configuration of the proposed joint emotion predictor, we adopted the HyperClova~\cite{kim2021changes}, which is a Korean variant of GPT-3 trained on a Korean-centric text corpus, as a feature extraction network for predicting emotional attributes.
In detail, the model consisted of 12 transformer decoder layers~\cite{vaswani2017attention} with 768 hidden units and 16 attention heads and was trained by using a large-scale text corpus consisting of 560 billion tokens.
The resulting network parameter was in a total of 137 million\footnote{
    This model corresponded to the smallest type in the original HyperClova paper.
    Noted that the model size was relatively smaller compared to the largest HyperClova with 82 billion parameters~\cite{kim2021changes}.
    In our preliminary experiments, however, we confirmed that 137 million parameters were enough to effectively train the emotion predictor while maintaining a low computational cost.
}.
For the two FC layers, one hidden layer with 256 hidden units followed by ReLU activation was used.
Each FC layer had softmax and linear output layers with the dimensions of 4 and 1, respectively.

To configure the joint emotion encoder, we used the LUT to convert the predicted emotion class to one of four feature vectors with a 32-dimensional size.
The input and output dimensions of projection matrix were set to 32.

\subsubsection{TTS model}
\label{sec:tts_model}
For the acoustic model, the duration predictor consisted of three FC layers and \oedit{a} uni-directional long short-term memory (LSTM) layer with 256 units. 
Each FC layer had 1024, 1024, and 512 units, respectively.
The Tacotron 2 encoder consisted of three convolution layers with 10$\times$1 kernels and 512 channels, bidirectional LSTM with 512 memory blocks, and FC layers with 512 units.
The Tacotron 2 decoder was composed of PreNet, PostNet, and the main LSTM block.
The PreNet consisted of two FC layers with 256 units.
The main LSTM block consisted of the two unidirectional LSTM layers with 1,024 memory blocks, followed by two FC layers with 79 units.
Finally, the PostNet consisted of five convolution layers with 5$\times$1 kernels and 512 channels, followed by a residual connection to its input.

For the configuration of the MB-HN-PWG vocoder, the harmonic WaveNet consisted of 20 dilated residual blocks with two exponentially increasing dilation cycles, whereas the noise WaveNet consisted of 10 residual blocks with one exponentially increasing dilation cycle.
The number of residual and skip channels was set to 64, and the convolution filter size was five.
More setup details for the neural vocoder can be found in our previous work \cite{hwang2021high}.

\begin{figure}[!t]
	\centering
	\begin{subfigure}[b]{0.237\textwidth}
	    \centering
	    \includegraphics[width=\linewidth]{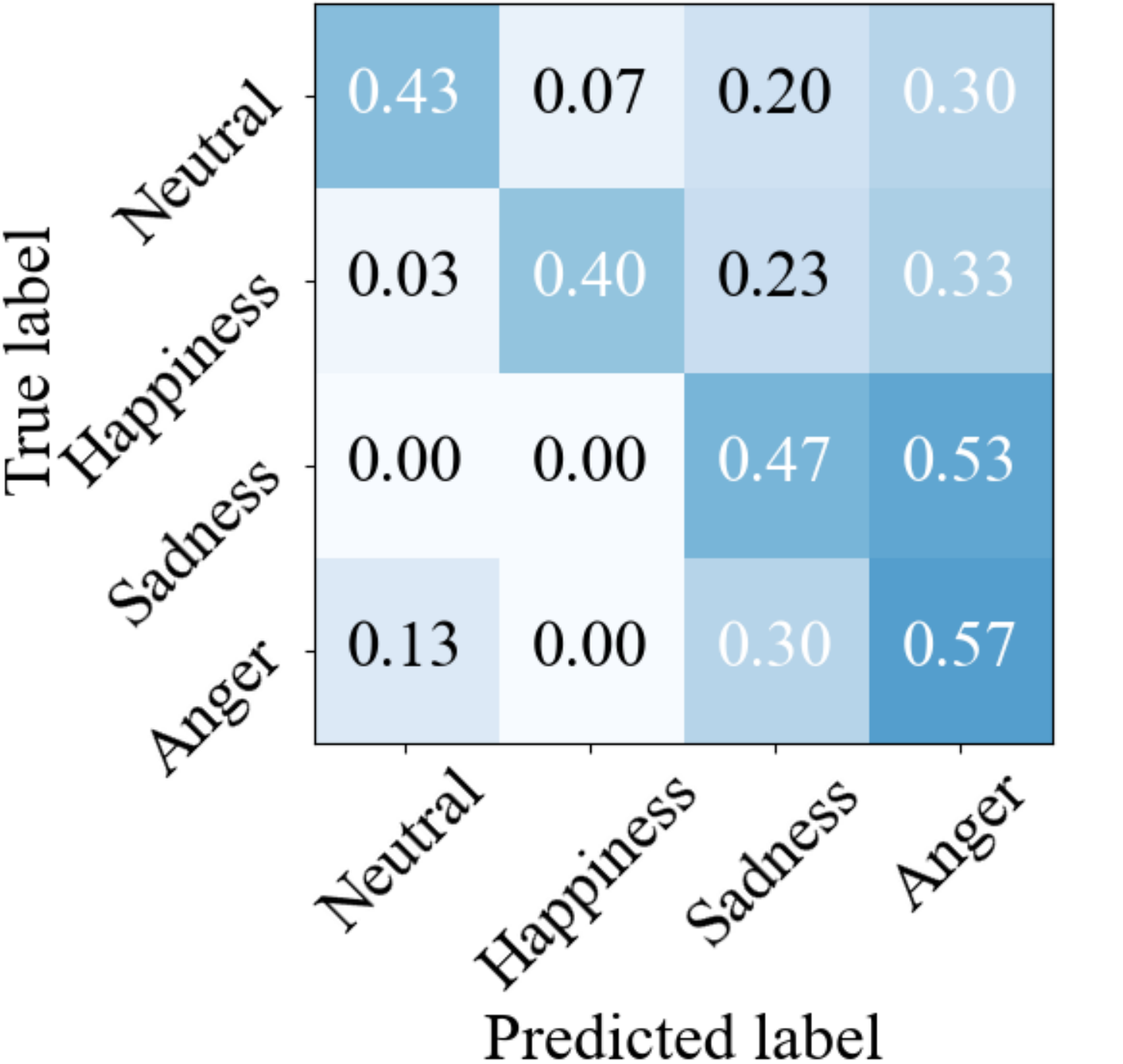}
   	    \caption{}
	\end{subfigure}
	\begin{subfigure}[b]{0.223\textwidth}
	    \centering
	    \includegraphics[width=\linewidth]{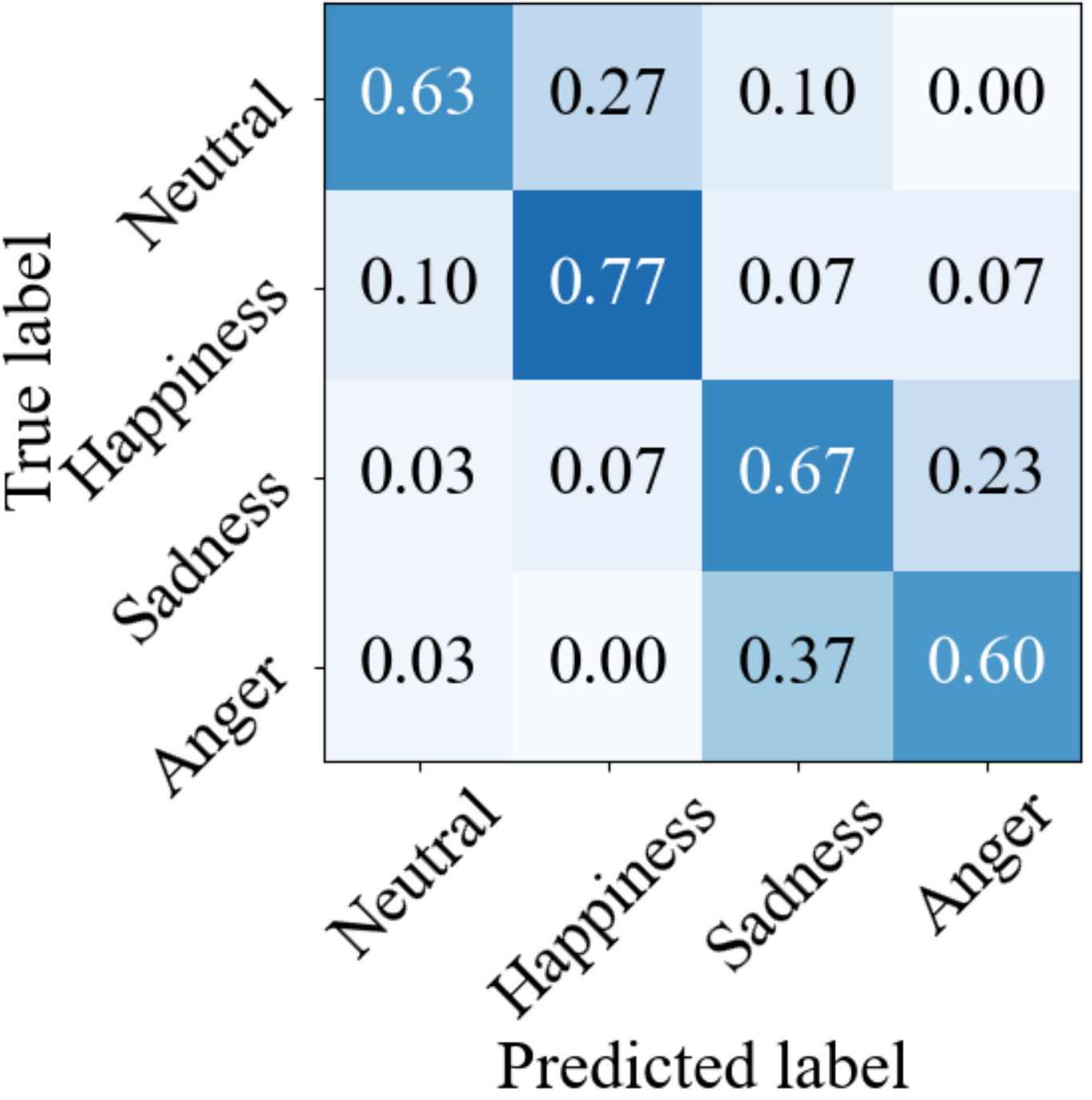}
  	    \caption{}
	\end{subfigure}
	\caption{
	    Classification accuracy of emotion predictors: (a) conventional CNN-based and (b) proposed LM-based methods
	}
	\label{fig:acc_confusion_matrix}
	\vspace*{-6pt}
\end{figure}

\subsection{Evaluations}
\subsubsection{Effectiveness of LM for emotion prediction}
\label{sec:exp_classification}
To verify the advantages of \yedit{the} LM-based emotion prediction, we compared the classification error of the emotion class predicted by conventional CNN-based and LM-based emotion predictors.
As shown in Figure~\ref{fig:acc_confusion_matrix}, our proposed emotion predictor achieved significantly higher classification accuracy than the conventional CNN-based method.
For instance, the proposed LM-based emotion predictor scored 76.7\% on classification accuracy when predicting happiness emotion, which is 36.7\% higher than that of the conventional one.
Overall, the classification accuracy of the proposed emotion predictor averaged by all emotions was 66.7\%, which was 20.0\% higher than the conventional one's 46.7\% prediction accuracy.

\begin{table*}[]
\centering
\footnotesize
\caption{
    The naturalness MOS test results with a 95\% confidence interval.
    Note that in the test, the listeners were asked to evaluate the quality of each audio sample.
}
\setlength\tabcolsep{4pt}
\renewcommand{\arraystretch}{0.7}
\hspace{-2mm}
\label{tab:mos_score_1}
\begin{tabular}{cccccccccc}
\Xhline{2\arrayrulewidth}
\begin{tabular}[c]{@{}c@{}}System\\ Index\end{tabular} & \begin{tabular}[c]{@{}c@{}}Type of\\ Emotion predictor\end{tabular} & \begin{tabular}[c]{@{}c@{}}Use of joint\\ Emotion encoder\end{tabular} & \begin{tabular}[c]{@{}c@{}}Emotion\\ Class\end{tabular} & \begin{tabular}[c]{@{}c@{}}Emotion\\ Strength\end{tabular} & Neutral & Happiness & Sadness & Anger & Average \\ \Xhline{2\arrayrulewidth}
S1                                                     & --                                                                  & --                                                                     & Manual                                                  & --                                                         &    4.12$\pm$0.22  &  4.20$\pm$0.24   &  3.85$\pm$0.30  &  3.46$\pm$0.23  &     3.90$\pm$0.17\\
S2                                                     & --                                                                  & Yes                                                                    & Manual                                                  & Manual                                                     &         4.17$\pm$0.19        &      4.23$\pm$0.21               &       3.83$\pm$0.28              &  3.59$\pm$0.21     &  3.96$\pm$0.15         \\
S3                                                     & CNN~\cite{cui2021emovie}                                                                 & --                                                                     & Predicted                                               & --                                                         &         3.96$\pm$0.15          &     3.75$\pm$0.24                 &     3.91$\pm$0.23                 & 3.56$\pm$0.15   &     3.79$\pm$0.09         \\
S4 (Proposed)                                             & LM~\cite{kim2021changes}                                                                  & Yes                                                                    & Predicted                                               & Predicted                                                  &         4.12$\pm$0.22        &     4.19$\pm$0.24        &      3.84$\pm$0.32                &      3.52$\pm$0.23                &   3.92$\pm$0.15         \\ \hline
Recordings                                             & --                                                                  & --                                                                     & --                                                      & --                                                         &         4.64$\pm$0.17 & 4.29$\pm$0.30 & 4.63$\pm$0.20 & 4.69$\pm$0.20 & 4.56$\pm$0.09         \\ \Xhline{2\arrayrulewidth}
\end{tabular}
\end{table*}

\begin{table*}[]
\centering
\footnotesize
\caption{
    Emotion\yedit{al} expressiveness MOS test results with a 95\% confidence interval.
    Note that in the test, the listeners were asked to evaluate how the emotions of synthetic speech matched the content of the text.
}
\setlength\tabcolsep{4pt}
\renewcommand{\arraystretch}{0.7}
\label{tab:mos_score_2}
\begin{tabular}{cccccccccc}
\Xhline{2\arrayrulewidth}
\begin{tabular}[c]{@{}c@{}}System\\ Index\end{tabular} & \begin{tabular}[c]{@{}c@{}}Type of\\ Emotion predictor\end{tabular} & \begin{tabular}[c]{@{}c@{}}Use of joint\\ Emotion encoder\end{tabular} & \begin{tabular}[c]{@{}c@{}}Emotion\\ Class\end{tabular} & \begin{tabular}[c]{@{}c@{}}Emotion\\ Strength\end{tabular} & Neutral & Happiness & Sadness & Anger & Average \\ \Xhline{2\arrayrulewidth}
S1                                                     & --                                                                  & --                                                                     & Manual                                                  & --                                                         &    4.00$\pm$0.24          &     4.07$\pm$0.23                 &       3.87$\pm$0.32               &  3.61$\pm$0.33  &       3.89$\pm$0.10\\
S2                                                     & --                                                                  & Yes                                                                    & Manual                                                  & Manual                                                     &         4.09$\pm$0.23        &        4.14$\pm$0.23              &          3.86$\pm$0.25            &   3.77$\pm$0.29   &      3.96$\pm$0.09         \\
S3                                                     & CNN~\cite{cui2021emovie}                                                                 & --                                                                     & Predicted                                               & --                                                         &         3.64$\pm$0.15         &      2.52$\pm$0.21                &      3.34$\pm$0.16                &  3.53$\pm$0.24  &    3.26$\pm$0.25         \\
S4 (Proposed)                                             & LM~\cite{kim2021changes}                                                                  & Yes                                                                    & Predicted                                               & Predicted                                                  &         4.22$\pm$0.25       &      4.01$\pm$0.28       &       3.84$\pm$0.34              &        3.67$\pm$0.30              &   3.94$\pm$0.12         \\ \hline
Recordings                                             & --                                                                  & --                                                                     & --                                                      & --                                                         &         4.83$\pm$0.13 & 4.46$\pm$0.32 & 4.69$\pm$0.21 & 4.65$\pm$0.20 & 4.66$\pm$0.08         \\ \Xhline{2\arrayrulewidth}
\end{tabular}
\end{table*}

\begin{table}[]
\centering
\footnotesize
\caption{
    Subjective preference test results (\%) between two different types of generation cases.
    The systems that achieved significantly better preference at the \textit{p} < 0.01 level are in boldface.
}
\renewcommand{\arraystretch}{0.7}
\hspace{-2mm}
\begin{tabular}{lccccc}
\Xhline{2\arrayrulewidth}
Index & S2  &  S4\textsubscript{single} & S4\textsubscript{multi} & Neutral & $p$-value \\ \hline
Test 1    & 40.6    & --  & 40.0   & 19.4  & 0.47\\ 
\textbf{Test 2}    & \textbf{--}       & \textbf{24.4} & \textbf{55.0} &  \textbf{20.6} & $\mathbf{<10^{-3}}$\\ 	
\Xhline{2\arrayrulewidth}
\end{tabular}
\label{tab:ab_test}
\begin{tabular}{l}
\end{tabular}
\vspace*{-3mm}
\end{table}

\subsubsection{Subjective listening tests}
For the subjective evaluation, all TTS systems had the same structures as \yedit{the} acoustic model and \yedit{the} neural vocoder, as described in Section~\ref{sec:tts_model}, except for the method \yedit{of} deciding emotional attributes.
First, we evaluated the system without (S1) or with (S2) the proposed joint emotion encoder by using manually annotated emotional attributes as described in Section~\ref{sec:annotation}.
We then evaluated the \yedit{accuracy of} predicting emotional attributes by using conventional CNN-based (S3) and proposed LM-based (S4) emotion predictors.
By following the method of Cui et al.~\cite{cui2021emovie} described in Section~\ref{sec:emotion_predictor}, the CNN-based emotion predictor estimated emotion class only.
Noted that when the emotion encoder was not used, a simple LUT layer was applied to \oedit{the} emotion class to compute emotion embedding.

We conducted two MOS listening tests to evaluate the (1) naturalness and (2) emotional expressiveness of synthetic speech.\footnote{
    Generated audio samples are available at the following URL:\\ \url{https://christophyoon.github.io/lmemotiontts/}}
During the tests, \oedit{a total of} 18 native Korean listeners were asked to rate the score of speech samples using the following 5-point responses: 1 = Bad, 2 = Poor, 3 = Fair, 4 = Good, and 5 = Excellent.
We randomly selected 15 sentences per emotion from the test set and then synthesized speech signals using the different models; thus, a total of 1,080 scores were collected to evaluate each system.

For the naturalness MOS test, the listeners were asked to rate how close the perceptual quality of the synthetic speech was to that of the recordings.
The trends of the test results in Table~\ref{tab:mos_score_1} can be analyzed as follows:
(1) Utilizing the emotion strength by using the proposed joint emotion encoder in addition to the emotion class improved the perceptual quality of the emotional TTS system (S1 vs. S2).
(2) When the emotional attributes were predicted from text, the TTS system with the proposed LM-based emotion predictor provided a significantly higher quality of synthetic speech than the system with conventional CNN-based one (S3 vs. S4).
(3) When the emotional attributes were predicted from the text, there was \yedit{a} quality degradation due to the prediction error of the emotion predictor (S1 vs. S3 and S2 vs. S4).
(4) However, the TTS system with the proposed LM-based emotion predictor showed a competitive quality of synthetic speech to the system with manually decided emotional attributes by achieving only a 0.04 lower MOS result on average (S2 vs. S4).

For the emotion expressiveness MOS test, listeners were asked to score how the emotion of synthetic speech matched the context of the input text.
The trends of the test results in Table~\ref{tab:mos_score_2} can be analyzed as follows:
(1) The usage of both emotion class and strength improved the emotional expressiveness compared to the system using only emotion class (S1 vs. S2).
This also verified the advantages of the proposed emotion encoder for better emotion expressiveness.
(2) When the emotional attributes were predicted through the emotion predictor, there was \yedit{a} quality degradation in terms of emotion expressiveness (S1 vs. S3 and S2 vs. S4).
This was mainly caused by misclassification in emotion class, which often generated utterances that did not match the emotional context.
(3) Nonetheless, the proposed method still outperformed the conventional method significantly (S3 vs. S4).
In particular, there was a large gap in the expressiveness score of happiness, where the classification error gap between conventional and proposed methods was the largest as explained in Section~\ref{sec:exp_classification}.
This indicated that a more accurate prediction reduced the peculiarity that occurred in the \yedit{emotion misclassification}.
(4) Finally, the gap between the LM-based emotion predictor and the manually decided system was only 0.02 on average (S2 vs. S4), indicating that the predicted emotional attributes were reliable.

\subsubsection{Paragraph-level emotional speech synthesis}
\label{sec:para-level}
The merits of the proposed emotion prediction method could also be found in paragraph-level speech synthesis because the LM had the capacity to capture the emotional context among the multiple sentences.
To verify this, we conducted \yedit{two} A-B preference tests by using various audiobook scripts.
In \yedit{these} tests, 18 native Korean listeners were asked to select a more preferable one between two generated speech samples.
We randomly selected 10 paragraphs \oedit{consisting} of 6--7 sentences, and then synthesized corresponding speech corpora using three different synthesis methods as follows:
\begin{itemize}[leftmargin=*]
    \item S2: The emotional TTS system that \textit{manually} annotated the emotional attributes.
    In particular, the class and strength were decided by considering the context of the input.
    \item S4\textsubscript{single}: The emotional TTS system with the proposed emotion predictor that received only a single utterance as an input at a time.
    Note that this model did not consider emotional context among consecutive sentences.
    \item S4\textsubscript{multi}: The emotional TTS system with the proposed emotion predictor that received \textit{multiple utterances} at once.
    In this model, the context of consecutive sentences was considered for predicting emotional attributes.
\end{itemize}

Table~\ref{tab:ab_test} shows the results of the preference tests, whose trend can be summarized as follows:
(1) The TTS system with proposed emotion modeling methods provided statistically indistinguishable synthetic speech compared to the TTS system using manually annotated emotional attributes (Test 1; S2 vs. S4\textsubscript{multi}).
This indicated that the proposed emotion prediction method had the capability of replacing the human-annotating process.
(2) The listeners preferred using the multiple sentences to using only a single sentence for the input of the emotion predictor (Test 2; S4\textsubscript{single} vs. S4\textsubscript{multi}).
This verified that the usage of LM was effective in generating paragraph-level content requiring consideration of the input sentences' context.

\section{Conclusion}
We proposed LM-based effective emotion prediction methods for an emotional TTS system.
As our method employed \yedit{the} LM to predict the emotion class and strength from \eedit{the} text \eedit{input}, the proposed \eedit{system could generate the corresponding emotional voice} without relying on auxiliary inputs.
The experimental results verified that the proposed method \eedit{provided} \eedit{a superior quality over the conventional methods and} a comparable quality to a system requiring \eedit{human} labeling.
Future works include \yedit{utilizing} more complex emotional representation method such as global style token \cite{stanton2018predicting} or variational auto-encoder \cite{kingma2013auto} that can further improve the expressiveness of the synthetic speech.

\section{Acknowledgements}
This work was supported by Clova Voice, NAVER Corp., Seongnam, Korea.

\newpage

\bibliographystyle{IEEEtran}
\bibliography{mybib}

\end{document}